\documentclass[nopreprint,twocolumn]{jasatex}
\usepackage{bm}

\begin{document}

\title{
Dynamical Energy Analysis - determining wave energy 
distributions in complex vibro-acoustical structures
}
\author{Gregor Tanner}
\email{gregor.tanner@nottingham.ac.uk}
\affiliation{School of Mathematical Sciences, University of Nottingham, University Park, 
Nottingham NG7 2RD, UK}
\begin{abstract}
We propose a new approach towards determining the distribution of mechanical 
and acoustic wave energy in complex built-up structures. 
The technique interpolates between standard {\em Statistical Energy Analysis} 
(SEA) and full {\em ray tracing} containing both these methods as limiting case.  
By writing the flow of ray trajectories in terms of linear phase space operators,
it is suggested here to reformulate ray-tracing algorithms in terms of 
boundary operators containing only short ray segments. 
SEA can now be identified as a low resolution ray tracing algorithm and typical
SEA assumptions can be quantified in terms of the properties of the ray 
dynamics.
The new technique presented here enhances the range of applicability of 
standard SEA considerably by systematically incorporating dynamical 
correlations wherever necessary.
Some of the inefficiencies inherent in typical ray tracing 
methods can be avoided using only a limited amount of the geometrical
ray information. 
The new dynamical theory - {\em Dynamical Energy Analysis} (DEA) - 
thus provides a universal approach towards determining wave energy 
distributions in complex structures. 
\end{abstract}
\date{\today}
\pacs{}
\maketitle
\section{Introduction}
\label{sec;intro}
Wave energy distributions in complex mechanical systems can often be modelled 
well by using a thermodynamical approach. Lyon 
argued as early as 1967 \cite{Lyo67} that the flow of wave energy follows 
the gradient of the energy density just like heat energy flows along 
the temperature gradient. To simplify the treatment, it is often 
suggested to partition the full system into subsystems and to assume 
that each subsystem is internally in 'thermal' equilibrium. Interactions 
between directly coupled subsystems can then be described in terms of 
coupling constants determined by the properties of the wave dynamics at 
the boundaries of intersection alone. These ideas formed the basis of 
{\em Statistical Energy Analysis} (SEA) which has since 
become an important tool in mechanical engineering and has been described 
in detail in text books such as  by Lyon and DeJong \cite{LD95}, Keane and 
Price \cite{KP94} and Craik \cite{Cra96}.  

A method similar in spirit but very different in its applications is the 
so-called {\em ray tracing technique}. The wave intensity distribution at
a specific point $r$ is determined here by summing over contributions
from ray paths starting at a a source and reaching the receiver point $r$ 
and thus by the flow of ray trajectories.  The method has found widespread 
applications in room acoustics \cite{Kut00} and seismology \cite{Cer01} 
as well as in determining radio wave field distributions in wireless 
communication \cite{MH91} and in computer imagining software \cite{Gla89}. 
A discussion of ray tracing algorithms used for analysing the energy 
distribution in vibrating plates can be found in \cite{CI01}. 

Both methods - that is, SEA and ray tracing -
are in fact complementary in many ways. Ray tracing can handle 
wave problems well, in which the effective number of reflections 
at walls or interfaces is relatively small. It gives estimates for the 
wave energy density with detailed spatial resolution and works for all 
types of geometries and interfaces. SEA can deal with complex structures 
carrying wave energy over many sub-elements including potentially a large 
number of reflection and scattering events albeit at the cost of reduced 
resolution. In addition, the quality of SEA predictions may depend on how
the subsystems are chosen as well as on the geometry of 
the subsystems itself, and a priory error bounds are often hard to obtain.

Ray tracing and SEA have in common that they predict mean values
of the energy distribution and do not contain information about wave effects
such as interference, diffraction or tunnelling giving rise to 
short scale modulations of the signal on the scale of a wave length.
Both methods are thus expected to hold 
in the high frequency or small wave length limit where the 
small scale fluctuations in the wave solutions are often averaged 
out, for example, due to a finite resolution of the receiver.

It will be shown here that SEA can be derived from a ray picture and 
is indeed a low resolution version of a ray tracing method. This makes it 
possible to introduce a new technique which interpolates between SEA and 
a full ray tracing analysis. The new method - 
{\em Dynamical Energy Analysis} (DEA) - keeps as much information 
about the underlying ray dynamics as necessary, benefiting at the same time 
from the simplicity of the SEA ansatz. DEA is thus an SEA type 
method in spirit but enhances the range of applicability of standard 
SEA considerably and makes it possible to give quantitative error bounds for
an SEA treatment. 

The ideas as presented here have their origin in wave or quantum chaos theory 
in which short wave length approximations are combined with dynamical 
systems or chaos theory, see \cite{TS07a} for an overview. Methods similar in 
spirit to the 
theory outlined in this paper have been discussed in the context of 
structural dynamics before.  Heron \cite{Her94} modelled 
correlations between energy densities in subsystems which are not
adjacent to each other in terms of direct and indirect contributions; 
the method does not take into account the actual ray dynamics and thus 
neglects long range dynamical correlations. Langley's \cite{Lan92} {\em 
Wave Intensity Analysis} (WIA) treats the wave field within 
each sub-component as an (inhomogeneous) superposition of plane 
waves thus introducing directionality which can propagate across 
coupling boundaries. The wave field is, however, assumed to be spatially 
homogeneous in each subsystem - an ad-hoc assumption which may often not 
be fulfilled.  A ray-tracing treatment developed in a series of papers 
by Le Bot \cite{LB98} is probably closest to the approach presented here; by employing
local power balance equations, a Green function for the mean energy flow
is obtained and the full flow across subsystems is obtained via 
flux conditions. The approach differs in as far as we consider 
multi-reflection in terms of linear operators here directly and use a 
basis function representation of these operators leading to 
SEA-type of equations.\\

The paper is structured as follows: in Sec.\ \ref{sec:SEA}, we will briefly 
review the ideas behind standard SEA. In Sec.\ \ref{sec:Green} the ray tracing 
approximation will be derived starting  from the Green function and 
using small wave length asymptotics. In 
Sec.\ \ref{sec:raydyn}, we will introduce the concept of phase space 
operators and their representation in terms of boundary basis functions. 
SEA emerges when restricting the basis set to constant functions only. A 
specific example - coupled two plate system will be treated in Sec.\ 
\ref{sec:2plates}.

\section{SEA revisited}
\label{sec:SEA}
Starting point for an SEA treatment is the division of the whole system into
subsystems; this will be done usually along natural boundaries, such as joints
between plates or walls in a building.  Energy is pumped into the system
at localised or delocalised source points (such as the vibrations of a motor) 
and is distributed throughout the systems in terms of vibrational or acoustic 
energy in one form or another. The net power flow between subsystems is then 
given in the simple form
\begin{equation} \label{SEA}
P_{ij} = \omega n_i \eta_{ij} \left(\frac{E_i}{n_i} - 
\frac{E_j}{n_j}\right)\, ,
\end{equation}
where $P_{ij}$ is the power transmitted between subsystem $j$ and $i$, 
$\omega$ is the (mean) frequency of the source, $n_i$ is the modal density 
of the (uncoupled) subsystem $i$, $\eta_{ij}$ is a coupling
constant and $E_i$ is the total vibrational energy in subsystem $i$. 
Allowing for a source term and dissipation and getting estimates for the 
coupling constants \cite{LD95,MS00,Mac03} and the modal densities via 
Weyl's law, one obtains a linear systems of equations which is solved for 
the unknown energies $E_i$. SEA gives mean values for these energies in the 
same way as Weyl's law gives the mean density of eigenfrequencies. 

The validity of Eq.\ (\ref{SEA}) is based on a set of conditions. 
It is assumed that subsystems have no memory, that is, the coupling constants
$\eta_{ij}$ depend on the properties of subsystems $i$ and $j$ alone. 
The eigenfunctions of the (uncoupled) subsystems are furthermore
expected to be locally 
described in terms of random Gaussian fields (''diffusive wave fields''). 
These key assumptions are expected to be valid only in the high frequency 
regime \cite{Mac03} and for weakly coupled subsystems \cite{Lan90}.

By the nature of the technique, only relatively rough estimates for energy
distributions can be obtained. Still, for high frequency noise sources,
SEA or variants thereof are often the method of choice. `Exact' solution
tools such as finite element methods (FEM) become both too
expensive computationally and unreliable, that is, small uncertainties in
the systems may lead to very different outputs.
One of the big challenges in mechanical engineering is the so-called
{\em mid-frequency problem} - that is, handling the frequency range which is
out of reach for `exact' numerical methods but not yet in the
high-frequency regime where SEA or ray methods are expected to work. 
SEA has been used as a starting point for penetrating the mid-frequency 
regime by employing hybrid-methods based on combining 
FEM and SEA treatments \cite{Fre97,LB99,MS00,SL05}.

Connections between SEA and the properties of a ray dynamics associated 
with the wave equation has so far been made only indirectly. 
The statistical properties of wave systems with a chaotic classical ray 
dynamics have been shown to follow random matrix theory with wave functions
behaving like random Gaussian waves \cite{GMW98}. The basic SEA assumptions 
thus imply that the ray dynamics in each sub-element needs to be 
chaotic.  This point of view has been stressed in the 
SEA literature by Weaver \cite{Wea89} and more recently in the context of 
determining the variance of the wave output data in \cite{LB04,LC04,CLK05}. 
A detailed review discussing the connections between ray and wave chaos has 
been given by Tanner and S\o ndergaard \cite{TS07a}.

\section{Wave energy density - from the Green function to the diagonal 
approximation}
\label{sec:Green}
\subsection{The Green function}
We assume that the system as a whole is characterised  by a linear wave 
operator $\hat{H}$ describing the overall wave dynamics, that is, the 
motion of all coupled subcomponents 
as well as damping and radiation. Different types of wave equations may be 
used in different parts of the system typically ranging from the 
Helmholtz equation for thin membranes and acoustic radiation to the 
biharmonic equation for plate-like elements and to vector wave equations 
describing in-plane modes in plates and bulk elasticity in isotropic or 
anisotropic media.  We restrict the treatment here to stationary problems 
with continuous, monochromatic energy sources - generalising the results to 
the time domain with impulsive sources is straight forward. 

To simplify the notation, we 
will in the following assume that $\hat{H}$ is a scalar operator; treating
bulk elasticity does not pose conceptual problems and 
follows for isotropic problems from Ref.\ \cite{TS07b} and for the anisotropic 
case, for example, from Ref.\ \cite{Son07}. Note that both in SEA as in the 
new method - DEA- developed below, different wave modes such as pressure, 
shear or bending waves will be treated as different subsystems.

The general problem of determining the response of a system to external 
forcing can then be reduced to solving
\begin{equation} \label{weqn}
\left(\omega^2 - \hat{H}\right) G(r,r_0,\omega) = \frac{F_0}{\rho}\, \delta(r -r_0)
\end{equation}
where the Green function $G(r,r_0,\omega)$ represents the wave amplitude 
induced by a force $F_0$ (of unit strength) acting continuously at a 
source point $r_0$ with driving frequency $\omega$; $\rho(r)$ denotes the 
material density. The wave energy density induced by the source is
\begin{equation} \label{wdens}
\epsilon_{r_0}(r,\omega) \propto \rho \omega^2 |G(r,r_0,\omega)|^2 .
\end{equation}

The bulk of the literature in acoustics and vibrational dynamics continues at 
that point by expanding the Green function in terms of eigenfunctions of 
either the full system or its sub-components. We propose to follow a 
different route here by introducing a connection between the energy 
density and an underlying ray dynamics and expressing the Green 
function in terms of classical rays. A brief overview introducing the 
Eikonal approximation and the notation used for describing the ray dynamics
is given in App.\ \ref{app:eikonal}.

\subsection{Small wave length asymptotics of the Green function}
\label{sec:asym}
Using small wave length asymptotics, the Green function $G({ r}, {r_0}, 
\omega)$ can be written as sum over \underline{all} 
classical rays going from $r_0$ to $r$ for fixed $H(r,p) = \omega^2$, 
where $H$ is the Hamilton function associated with the operator
$\hat{H}$, such as (\ref{Hamiltonian}), and $p$ is the momentum variable
or wave number vector, see App.\  \ref{app:eikonal}. One obtains  \cite{Gut90}
\begin{equation} \label{Gsc}
G({r}, {r_0}, \omega) = C \sum_{j: r \to r_0} A_j e^{i S_j(\omega) -
i \mu_j \frac{\pi}{2}}
\end{equation}
with prefactor
\[ 
C = \frac{F_0}{\rho(r_0)} \frac{\pi}{\omega}\frac{1}{(2\pi{\rm i})^{(d+1)/2}},
\]
where $d$ is the space dimension. The action $S_j(\omega)$ is defined in 
Eq.\ (\ref{action}), and is usually the 
dominant $\omega$ dependent term. The amplitudes $A_j$ can be written 
in the form \cite{ST02,TS07b}
\begin{equation}
A_j = A_j^{(d)} A_j^{(c)} A_j^{(g)}\, 
\end{equation}
containing contributions due to damping (d), 
conversion and transmission/reflection coefficients (c) and geometrical 
factors (g). The damping factor is typically  of the form $A_j^{(d)} = 
\exp(-\alpha_j L_j)$ with $\alpha_j$, the damping rate and $L_j$, the 
geometric length of the trajectory. Furthermore, $A_j^{(c)}$ corresponds
to the product of reflection, transmission or mode conversion amplitudes
encountered by the trajectory $j$ at boundaries or material interfaces 
\cite{Blu96, BH98, Kut00,TS07a}. Finally, $A{(g)}$ contains 
geometric information and is of the form
\begin{equation} \label{ag}
\left|A^{(g)}\right|^2 = \frac{1}{|\dot{r}||\dot{r_0}|} 
\left|\frac{\partial^2 S}{\partial{r}^{\perp}
\partial{r}^{\perp}_0}\right| = \frac{1}{|\dot{r}||\dot{r_0}|}
\left|\frac{\partial p_0^{\perp}}{\partial{r}^{\perp}}\right| 
\end{equation}
where $|\cdot| = | \det(\cdot)|$ and the 
derivatives are taken in a local coordinate system ${r}^{\perp}, 
{r}_0^{\perp}$ perpendicular to the trajectory at the initial 
and final point.  The inverse of the Jacobian in Eq.\ (\ref{ag})
relates changes in the initial momentum perpendicular to the trajectory,  
${p}_0^{\perp}$, to changes in the final position ${r}^{\perp}$ 
on the 'energy' manifold $H = \omega^2 = const$.  The phase index
$\mu_j$ contains contributions from transmission/reflection coefficients
at interfaces and from caustics, that is, singularities in the 
amplitude in Eq.\ (\ref{ag}).

The representation(\ref{Gsc}) has been considered in detail in quantum 
mechanics, see the books by Gutzwiller \cite{Gut90}, St\"ockmann \cite{Sto99}
and Haake \cite{Haa01}. It is valid also for general wave equations in 
elasticity such as the biharmonic \cite{BH98} and the Navier-Cauchy equation
\cite{TS07b}; in the latter case, $G$ becomes matrix valued. While the 
approximation is based on a small wave length expansion, the sum over 
trajectories in
Eq.\ (\ref{Gsc}) often gives remarkably good results down to the mid and low 
frequency regime. Note that the summation in Eq.\ (\ref{Gsc}) is typically over 
infinitely many terms  where the number of contributing rays 
increase (in general) exponentially with the length of the trajectories 
included. This gives rise to  convergence issues, especially in 
the case of low or no damping, see \cite{TS07a} and references therein.\\

The wave energy density, Eq.\ (\ref{wdens}), can now be expressed as a 
double sum over classical trajectories, that is,
\begin{eqnarray} \label{diag}
\epsilon_{{ r}_0}({ r}, \omega) &\propto&  
\sum_{j,j':{ r}_0 \to {\bf r}}  
A_j A_{j'}\, e^{i(S_j - S_{j'} - (\mu_j - \mu_{j'})\frac{\pi}{2}) }\\
&=& \rho({r}, { r_0}, \omega) + \mbox{off-diagonal terms}\, .
\nonumber
\end{eqnarray}
The dominant contributions to the double sum arise from terms in which the 
phases cancel exactly; one thus splits the sum into a {\em diagonal part}
\begin{equation} \label{diagsum}
\rho({r}, {r_0},\omega) = \sum_{j:{r}_0 \to {r}}  
|A_j|^2  
\end{equation}
containing only pairs with $j = j'$ in Eq.\ (\ref{diag}) and an 
{\em off-diagonal part} containing the rest. The diagonal contribution
gives a smooth background signal, which is  here proportional to the 
energy density; the 
off-diagonal terms give rise to fluctuations on the scale of the wave length. 
The phases related to different trajectories are (largely) uncorrelated 
and the resulting net contributions to the off-diagonal part are in general 
small compared to the smooth part - especially when considering averaging 
over frequency intervals of a few wave numbers. (There are exceptions from 
this general rule; length correlations between certain subsets of 
orbits can lead to important off-diagonal contributions. Coherent 
back-scattering or action correlations between periodic rays 
which have been identified to explain the universality of random matrix 
statistics are examples thereof; see \cite{TS07a} for details). 

In what follows, we will focus on the diagonal part, that is, we
will show that neglecting off-diagonal terms is equivalent to 
the standard ray tracing approximation. We will show furthermore
that ray tracing can be written in terms of linear phase space 
operators and that SEA can be derived as an approximation of these
 operator. The connection between SEA 
and classical (thermodynamical) flow equations is thus  put on sound 
foundations and the validity of the basic SEA assumptions as outlined in 
Sec.\ \ref{sec:SEA} can be quantified.

\section{Propagation of phase space densities - from ray tracing to SEA}
\label{sec:raydyn}
\subsection{Phase space operators and probability densities}
We consider the situation of a source localised at a point $r_0$ emitting
waves continuously at a fixed frequency $\omega$. Standard ray tracing
techniques estimate the wave energy at a receiver point $r$ by 
determining the density of rays in $r$ starting initially in $r_0$ 
(within the constraint $H(r_0,p_0) = \omega^2$) and reaching $r$ 
after some unspecified time. This can be written in the form
\begin{eqnarray} \label{dens1}
\rho({r}, {r_0}, \omega) = &&\int_0^{\infty} d\tau \int dp  \int dX' \\
\nonumber &&
\,w(X',\tau) \, \delta(X - \varphi^\tau(X')) \, \rho_0(X';\omega)
\end{eqnarray}
where ${X} = ({p},{r})$ denotes a point ins {\em phase space} and 
the initial density
\begin{equation} \label{initial}
\rho_0(X';\omega) = \delta({r}' - {r}_0) \delta(\omega^2 - H({X}')), 
\end{equation}
is centred at the source point $r_0$. Furthermore, 
$X(\tau)= \varphi^\tau(X')$ is the phase space flow generated by 
equations of motion of the form (\ref{eof-ray}) with initial conditions 
$X(0) = X'$ and $\tau$ is the time introduced in Eq.\ (\ref{eof-ray}).
It can be shown that Eq.\ (\ref{dens1}) is equivalent to the diagonal
approximation, Eq.\ (\ref{diagsum}), see App.\ \ref{app:ray-trace}. 

The weight function $w(X,\tau)$ contains damping and reflection/transmission 
coefficients and we assume here that $w$ is multiplicative, that is,
\begin{equation} \label{multip}
 w(X, \tau_1) w(\varphi^{\tau_1}(X), \tau_2) = w(X, \tau_1+\tau_2),
\end{equation}
which is fulfilled for (standard) absorption mechanism and 
reflection processes. Note, that the integral kernel
\begin{equation}
{\cal L}^\tau(X,X') = w(X',\tau) \delta(X - \varphi^\tau(X')) 
\end{equation}
is a linear operator - often called the Perron-Frobenius operator - which
(after setting $w=1$) may be interpreted as a propagator for the Liouville 
equation describing the time evolution of phase space densities
\cite{DasBuch}
\[ \dot{\rho}(X)) = \{H(X),\rho(X)\}\]
(where $\{\cdot,\cdot\}$ denotes the Poisson brackets) with solution
\[ 
\rho(X,\tau) = {\cal L}^\tau[\rho_0] = \int dX' \delta(X - \varphi^\tau(X')) 
\rho_0(X') .
\]
Eq.\ (\ref{dens1}) can be simplified to
\begin{eqnarray} \label{dens2}
\rho({r}, {r_0}, \omega) = &&\int_0^{\infty} d\tau 
\int dp'\,w(p',r_0,\tau)\, \\ \nonumber
&& \delta(r - \varphi_r^\tau(p',r_0))\,
\delta(\omega^2 - H(p',r_0))
\end{eqnarray}
where $\varphi_r^\tau(X) = r(\tau)$ denotes the $r$-component of the flow 
vector.  Eq.\ (\ref{dens2}) 
is the starting point for a variety of ray tracing techniques 
popular in room acoustics \cite{Kut00}, seismology \cite{Cer01} and 
optics such as illumination problems as well as for visualisation 
techniques in computer graphics \cite{Gla89}. 

While the basic equation (\ref{dens2}) may seem 'obvious' from a 
ray geometrical point of view, we provide in 
Sec.\ \ref{sec:asym} and App.\ \ref{app:ray-trace}
a derivation from first principles starting from
the wave equation. For references in a quantum context, see 
\cite{GV79, Ric00}.  The connection between the ray tracing
densities and the double sum over ray trajectories, Eq.\ (\ref{diag}),  
may form the basis for including ''higher order'' wave effects contained in 
the off-diagonal part. In what follows, we will stay within the diagonal 
approximations and use properties of the linearity of the phase space 
operator $\cal L$ to unveil the connection between ray tracing methods and 
SEA. 
%
\subsection{Boundary maps and related operators}

We will for simplicity assume that the
wave problem is confined to a finite domain with a well defined boundary; 
we may, for example, consider the vibrations of (coupled) plates of finite 
size or acoustics/elastic problems within bodies of finite volume. The 
long time limit of the dynamics is then best described in terms of boundary 
maps, that is, one records only successive reflections of a ray trajectory 
at the boundary. We introduce a coordinate system on the boundary, 
$X_s = (s,p_s)$, where $s$ 
parameterises the boundary and $p_s$ denotes the momentum components 
tangential to the boundary at $s$; ($X_s$ is often referred to as Birkhoff 
coordinates). Phase space points $X=(r,p)$ on the boundary are mapped
onto $X_s$ by an invertible transformation ${\cal B}:  X \to (X_s,\omega)$ 
with $H(X) = \omega^2$.

\begin{figure*}
\includegraphics[width=0.9\textwidth]{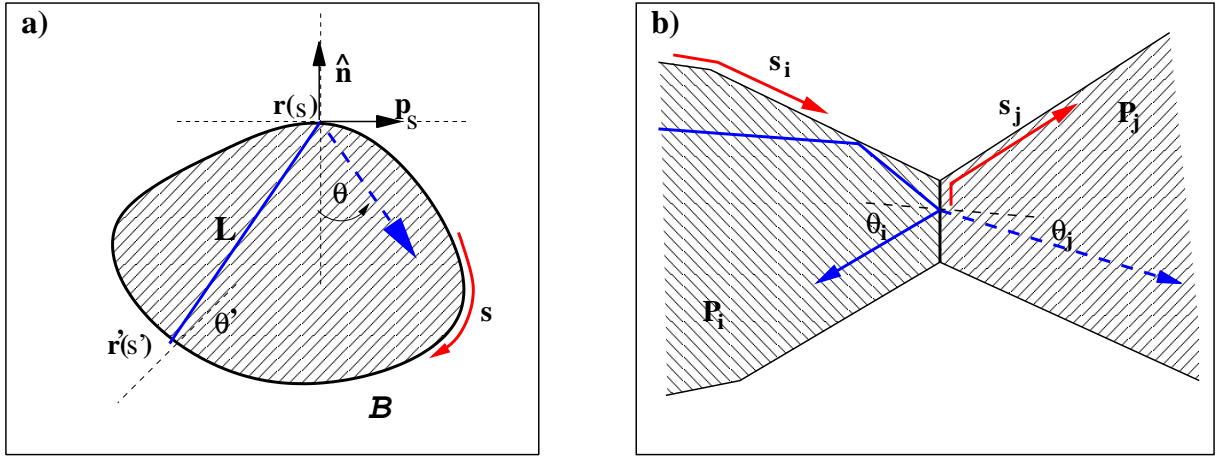}
\caption{Coordinates used for the boundary maps: 
a) in case of a single sub-system;
b) at an intersection between two sub-systems
\label{fig:coord}}
\end{figure*}

We now introduce two new operators: firstly, we define an operator 
${\cal L}_B$ propagating a source distribution from the interior to the 
boundary, that is, 
\[ {\cal L}_B(X_s, X') = w(X',\tau_B)\, \cos\theta \, 
\delta(X_s - {\cal B} (\varphi^{\tau_B}(X'))) \]
where $X'$ is an arbitrary phase space point in the interior and 
$\tau_B(X')$ is the time it takes for a trajectory with 
initial condition $X'$ to hit the boundary for the first time; the 
angle $\theta(X')$ is taken between the normal to the boundary at the 
point $s$ and the incoming ray velocity vector $p$, see 
Fig.\ \ref{fig:coord}a.  Secondly, we introduce the boundary operator 
\[ {\cal T}(X_s, X_s';\omega) = w(X'_s) \delta (X_s - \phi_\omega(X_s')), \]
which is the Perron-Frobenius operator for the boundary map
\[ 
\phi_\omega(X'_s) = {\cal B}(\varphi^{\tau_B}(X'))\quad \mbox{with} \quad
X' = {\cal B}^{-1}(X_s',\omega).
\]

One can now write the stationary density in the interior, Eq.\
(\ref{dens2}), in terms of the boundary operators introduced above. 
Firstly, the initial density (\ref{initial}) is mapped onto the boundary, 
that is, $\rho_0(X_s, \omega) = \int dX {\cal L}_B(X_s,X) \rho_0(X, \omega)$. 
The stationary density on the boundary induced by the source 
$\rho_0(X_s,\omega)$ is then
\begin{equation} \label{densb}
\rho(\omega) = \sum_{n=1}^\infty {\cal T}^n(\omega) \, \rho_0(\omega) = 
(1 - {\cal T}(\omega))^{-1} \rho_0(\omega)\, .
\end{equation}
where ${\cal T}^n$ contains trajectories undergoing $n$ reflections at 
boundary.  The resulting density distribution on the boundary, 
$\rho(X_s, \omega))$,
can now be mapped back into the interior using ${\cal L}_B^{-1}$ and 
one obtains the density (\ref{dens2}) after projecting down onto coordinate 
space, that is,
\begin{equation} \label{densi}
\rho(r,r_0, \omega) = \int dp\, dX_s\, {\cal L}_B^{-1}(X,X_s) 
\rho(X_s, \omega)\, . 
\end{equation}

The long term dynamics is thus contained in the operator $(1-{\cal T})^{-1}$
and standard properties of the Perron-Frobenius operators ensure that the 
sum over $n$ in Eq.\ (\ref{densb}) converges for non-vanishing dissipation. Note, 
that for $w(X) \equiv 1$, 
$\cal T$ has a largest eigenvalue 1 and the expression in Eq.\ (\ref{densb}) 
is singular. That is, in the case of no losses due to absorption or 
radiation, a source continuously emitting energy into the system will lead 
to a diverging energy density distribution in the 
large time limit. The eigenfunction of $\cal T$ (and $\cal L$) corresponding 
to the eigenvalue 1 is the constant function; that is, in equilibrium the 
energy is equally distributed over the full phase space \cite{Wea82}.

To evaluate $(1 - {\cal T})^{-1}$ it is convenient to express the 
operator $\cal T$ in a suitable set of basis functions defined on the 
boundary.  Depending on the topology of the boundary, complete function 
sets such  a Fourier basis for two dimensional plates or spherical 
harmonics for bodies in three dimensions may be chosen.  Denoting 
the orthonormal basis 
$\{\Psi_0, \Psi_1,\Psi_2, \ldots\}$, we obtain
\begin{eqnarray}\label{Tnm1}
T_{nm} &=& \int dX_s dX_s'\, \Psi^*_n(X_s)\, {\cal T}(X_s,X'_s;\omega)\, 
\Psi_m(X'_s) \\ \nonumber
&=& \int dX_s'\, \Psi^*_n(\phi_\omega(X'_s))\, w(X'_s)\, 
\Psi_m(X'_s) \, .
\end{eqnarray}
The treatment is reminiscent of the Fourier-mode approximation in the
wave intensity analysis (WIA) \cite{Lan92}; note, however, that the basis
functions cover both  momentum and position space and can thus
resolve space inhomogeneities unlike WIA.
If the boundary map $\phi_\omega(X_s)$ is not known or hard to obtain, it is
often convenient to write the operator in terms of trajectories
with fixed start and end point $s'$ and $s$; one obtains
\begin{eqnarray} \label{Tnm2}
T_{nm} &=& \int ds ds'\frac{1}{|\partial s/\partial p_s'|}\,
\Psi^*_n(X_s)\, w(X'_s) \,\Psi_m(X'_s)\\ \nonumber
&=& \int ds ds'
\left|\frac{\partial^2 S}{\partial s \partial s'}\right| \,
\Psi^*_n(X_s) \,w(X'_s)\, \Psi_m(X'_s)
\end{eqnarray}
with $X_s = (s,p_s(s,s'))$ and $X_s'= (s',p_s'(s,s'))$ and $S$ is the action 
introduced in Eq.\ (\ref{action}). The representation, Eq.\ (\ref{Tnm2}),
is advantageous for homogeneous problems where the ray 
trajectory connecting the points $s'$ and $s$ is a straight line, 
see the examples discussed in Sec.\ \ref{sec:2plates}.

\subsection{Subsystems}
In many applications, it is natural to split the full system into
subsystems and to consider the dynamics within each subsystem separately. 
Coupling between sub-elements can then be treated as losses in 
one subsystem and source terms in the other. Typical subsystem
boundaries are surfaces of reflection/transmission due to sudden changes 
in the material parameters or local boundary conditions due to for example
bends in plates. Also, weakly connected sub-domains such as two regions 
connected through small openings may be considered as separate subsystems.  
We denote the subsystems $\{P_1, \ldots P_N\}$
and describe the full dynamics in terms of the 
subsystem boundary operators ${\cal T}^{ij}$; flow from $P_j$ to $P_i$  
is possible only if the two subsystems are connected and one obtains
\begin{equation}  
{\cal T}^{ij}(X^i_s, X^j_s) = w^{ij}(X^j_s)\, 
\delta (X^i_s - \phi_\omega^{ij}(X^j_s))
\end{equation}  
where $\phi_\omega^{ij}$ is the boundary map in subsystem $j$ mapped onto the 
boundary of the adjacent subsystem $i$ and $X_s^i$ are the
coordinates of subsystem $i$, see Fig.\ \ref{fig:coord}b.  
(Note, that subsystems exchanging wave energy are necessarily connected 
through a common boundary here). The weight $w^{ij}$ contains, among 
other factors, reflection and transmission coefficients characterising the 
coupling at the interface between $P_j$ and $P_i$. 

A basis function representation of the full operator $T$ as suggested in Eq.\
(\ref{Tnm1}) is now written in terms of sub-system boundary basis 
functions $\Psi^i_n$ with 
\begin{equation}\label{Tnmsub}
T^{ij}_{nm} = 
\int\, dX^i_s dX^j_s \, {\Psi^i}^*_n(X^i_s)\, {\cal T}^{ij}(X^i_s,X^j_s) \,
\Psi^j_m(X^j_s).
\end{equation}
The equilibrium distribution on the boundaries of the subsystems is then 
obtained by solving the systems of equations (\ref{densb}) 
\begin{equation} \label{equb}
(1 - T) \rho = \rho_0\, .
\end{equation}
Here, $T$ is the full operator including all subsystems and the
equation is solved for the unknown energy densities 
$\rho = (\rho^1, \ldots \rho^N)$ where $\rho^i(n)$ denotes the 
(Fourier) coefficients of the density on the boundary of subsystem $i$.
Equations similar to (\ref{equb}) have been considered by Craik 
\cite{Cra96} in the context of SEA. Note, that for a source localised in 
subsystem $j$, one obtains $\rho^i_0 \ne 0$ only if $P_i$ has a 
boundary in common with subsystem $P_j$.

\subsection{From ray tracing to SEA}
Up to now, the various representations given in Sec.\ \ref{sec:raydyn}
are all equivalent and correspond to
a description of the wave dynamics in terms of the ray tracing ansatz
(\ref{dens1}). Traditional ray tracing based on sampling ray solutions over
the available phase space is rather inefficient, however. Convergence tends
to be fairly slow, especially if absorption is low and long paths including
multi-reflections need to be taken into account. Finding
all the possible rays which connect a fixed source and receiver point is a
computationally expensive boundary value problem and typically only a 
small sample of all the trajectories calculated are actually needed 
in determining the local energy density. In addition,
the number of rays connecting source and receiver grows quickly 
(often exponentially) with the length of the ray trajectories setting 
fairly tight numerical bounds on the number of reflections one can take 
into account - a severe limitation in the low damping regime. 

These problems are common for ray summation methods 
\cite{DasBuch, TS07a}. They can be overcome by describing the dynamics
in terms of boundary operators and  
boundary functions $\Psi_n$ as outlined above. While the representations 
are equivalent  when employing the full set of basis functions (leading 
to infinite dimensional operators $T$), this is, of course, not the case
for finite dimensional approximations. When considering the solutions 
of Eqs.\ (\ref{densb}) or (\ref{equb}), one is in general 
interested in smooth approximations of the energy density obtained 
from the classical flow. The resolution required is naturally limited 
by the wave length of the underlying wave equation, but in many 
application a much coarser resolution will be sufficient. Convergence 
for obtaining such coarse grained energy density distributions
is in general fast when increasing the dimension of the operators
involved and often only a very small number of basis functions 
(of the order $\le 10$ per subsystem and momentum and position coordinate) 
are necessary. In addition, only short ray-segments are needed 
to evaluate operators of individual subsystems as multi-reflections are 
included explicitly in the sum  (\ref{densb}). 

An SEA treatment emerges when approximating the individual operators  
${\cal T}^{ij}$ in terms of the lowest order basis function (or Fourier mode),
that is, the constant function $\Psi^j_0 = (A_B^j)^{-1/2}$ with $A^j_B$, 
the area of the boundary of $P_j$. The matrix $T^{ij}$ is then one-dimensional
and gives the mean transmission rate from subsystem $P_j$ to $P_i$. 
It is thus equivalent to the coupling loss factor $\eta_{ij}$ used in 
standard SEA equations (\ref{SEA}). The resulting full  
$N$-dimensional $T$ matrix (with $N$, the number of subsystems) 
yields a set of SEA equations using the relation  (\ref{equb})  
(after mapping the boundary densities back into 
the interior with the help of local operators ${\cal L}_B^i$). Note, that 
the terms $E_i/n_i \sim \bar{\rho}^i$ in Eq.\ (\ref{SEA}) are in fact mean 
energy densities as the mean density of eigenvalues is to leading order 
$n_i \propto A_i$ with $A_i$ the area/volume of subsystem $P_i$
following Weyl's law \cite{TS07a}. 

The matrix $T$ can in this approximation be interpreted as a transition matrix 
of an $N$ dimensional Markov chain; SEA is thus in fact a Markov approximation 
of a deterministic dynamics. Similar approaches have been taken in dynamical 
systems theory over the last decades leading to a stochastic 
interpretation of chaotic dynamical systems in terms of a thermodynamical 
formalism \cite{DasBuch}. A Markov or SEA approximation is thus justified if 
the ray dynamics within each subsystem is sufficiently chaotic such that a 
trajectory entering subsystem $j$ 'forgets' everything about its past 
history before exciting $P_j$ again.  In other words, correlations within 
the dynamics must decay fast on the timescales of the staying time 
$\bar{\tau}_j$, that is, on the time scale it takes for a typical ray to leave 
$P_j$ either by being transmitted to another subsystem $P_i$ or by being lost due
to absorption. The dynamics must indeed equilibrate on the 
time scale $\bar{\tau}_j$.
This condition will often be fulfilled if the subsystems' boundaries are
sufficiently irregular, the subsystems are dynamically well separated and 
absorption and dissipation is small - conditions typically cited in an SEA context.
In this case, SEA is an extremely efficient method compared to standard
ray tracing techniques.
However, for subsystems with regular features, such as rectangular cavities or
corridor-like elements, incoming rays are directly 
channelled into outgoing rays thus violating the equilibration hypothesis and
introducing memory effects. Likewise, strong damping may lead to significant
decay of the signal before reaching the exit channel introducing geometric
(system dependent) effects - that is, the distance between input and output
channel becomes relevant. \\

These features can all be incorporated by including higher order basis functions
for each subsystem boundary operator $T^{ij}$. This makes it possible 
to resolve the fine structure of the dynamics and its correlation as well as 
effects due to non-uniform damping over typical scales of the subsystem. 
As one increases the number of basis functions, a smooth interpolation from 
SEA to a full ray-tracing treatment is achieved. The maximal number of basis 
functions needed to reach convergence are expected to be relatively 
small thus making the new method more efficient than a full ray tracing 
treatment - in particular in the small damping regime. Typical dimensions 
of $T^{ij}$ are determined by escape-, correlation- and damping-rates 
of the ray dynamics in subsystem $j$. A priori or a posteriori bounds for 
the size of the basis set needed can thus be obtained from dynamical 
properties of the underlying ray flow. 

Representing the ray dynamics in terms of finite dimensional transition 
matrices may be regarded as a refined SEA technique. The new method takes 
advantage of the efficiencies of SEA, but includes additional information
about the ray dynamics where necessary, thus overcoming some of the 
limitations of SEA and putting the underlying SEA assumptions on sound
foundations. This gives rise to to the name  
{\em Dynamical Energy Analysis} (DEA). Note that, like SEA and ray 
tracing, the method is purely based on a classical ray picture and is thus 
inherently a short wavelength approximation. It does not take into account 
wave like phenomena; from a wave asymptotics point of view, these are 
contained in the off-diagonal contributions in Eq.\ (\ref{diag}). 
Wave effects often become important in mechanical structures containing
elements with short and long wave lengths (at the same basic frequency) 
and hybrid SEA - finite element methods have been developed in this case
\cite{LB99,MS00,SL05}. An extension of these methods to DEA will be of 
importance.

\section{A numerical example: a couple two-plate systems}
\label{sec:2plates}
The method has been implemented numerically for a coupled two-plate system;
the vibrational energy distribution has been calculated using DEA for 
plates of different shape where the coupling between the plates
is achieved by choosing simply supported boundary conditions (BC) along a 
common line of intersection. We assume clamped BC at the outer edges, that is, 
Snell's law of reflection applies and no losses occur at the boundaries.
The two plates have the same thickness and are homogeneous otherwise. 
The BC at the intersection introduces reflection and transmission and
acts as a barrier thus providing a natural boundary for dividing the system 
into two distinct subsystems. Some of the configurations considered are 
shown in the insets of Fig.\ \ref{fig:plate}. Estimates for the vibrational 
energy induced by a point source in subsystem 1 will be obtain by using 
DEA and will be compared to standard SEA results. 

\subsection{Set-up}
The plates are treated as two-dimensional systems and a Fourier basis 
both in position and momentum space is thus an adequate choice for the 
set of basis functions, that is
\[ \Phi^i_{\bf n}(s,p_s) = 
\frac{1}{\sqrt{2 L_i}} e^{2 \pi i (n_1 s/L_i + n_2 p_s/2)} ,
\]
with  ${\bf n} = (n_1,n_2)$,  $n_1,n_2$ integers,
and $ s\in [0,L_i), p_s \in (-1,1)$, where $L_i$ is the length of the 
boundary and $i=1$ or 2. The wave number $|p| \propto \sqrt{\omega} $ is 
set equal to 1 here. Note, that the energy distribution is expected to be 
frequency independent as neither the ray paths nor the reflection 
coefficients at the ray splitting boundary depend on $k$. We also assume
for simplicity that the damping coefficient $\alpha$ is independent of 
the driving frequency $\omega$. The transmission probability
at the intersection of the two plates yields for simply supported BC
\[ w_t(\theta) = \frac{1}{2} \cos^2 \theta \]
with $\theta \in [-\pi/2,\pi/2]$, the angle between the incoming ray and the 
normal to the surface.

Given the start and end point $s', s$ on the boundary of either plate 1 or 2, 
the rays, their lengths and the angles of intersection (and thus the momentum 
components tangential to the boundary) can be obtained easily, and the 
integral representation of the boundary operator in the form (\ref{Tnm2}) 
will be used. Writing out the Jacobian $|\partial s/\partial p'|$, one obtains 
\[
T^{ij}_{\bf nm} =\int ds^i ds^j w^{ij} \frac{\cos \theta^i \cos \theta^j}
{L(s^i,s^j)}
{\Phi^i}^*_{\bf n} \Phi^j_{\bf m}
\]
where $L(s^i,s^j)$ is the length of the trajectory. The weight function
is given as 
\[ w^{ij} = w^{ij}_b\, e^{-\alpha L}  \]
with $\alpha$, the damping coefficient, and the reflection/transmission 
coefficients are 
\[
w_b^{ij}(s^i,s^j) = \left\{ \begin{array}{ccc}
                           \delta_{ij} & \mbox{if} & s^i \notin B^i_I\\
                           \delta_{ij}  + (-1)^\delta_{ij} 
                           w_t(\theta^i(s^i,s^j))& \mbox{if} & s^i \in B^i_I\,,
                         \end{array} \right .
\]                            
where $B_I^i$ denotes the part of the boundary in the coordinate system
 $s^i$ lying on the intersection of plate 1 and 2.

\subsection{Numerical results}

\begin{figure*}
\includegraphics[width=0.8\textwidth]{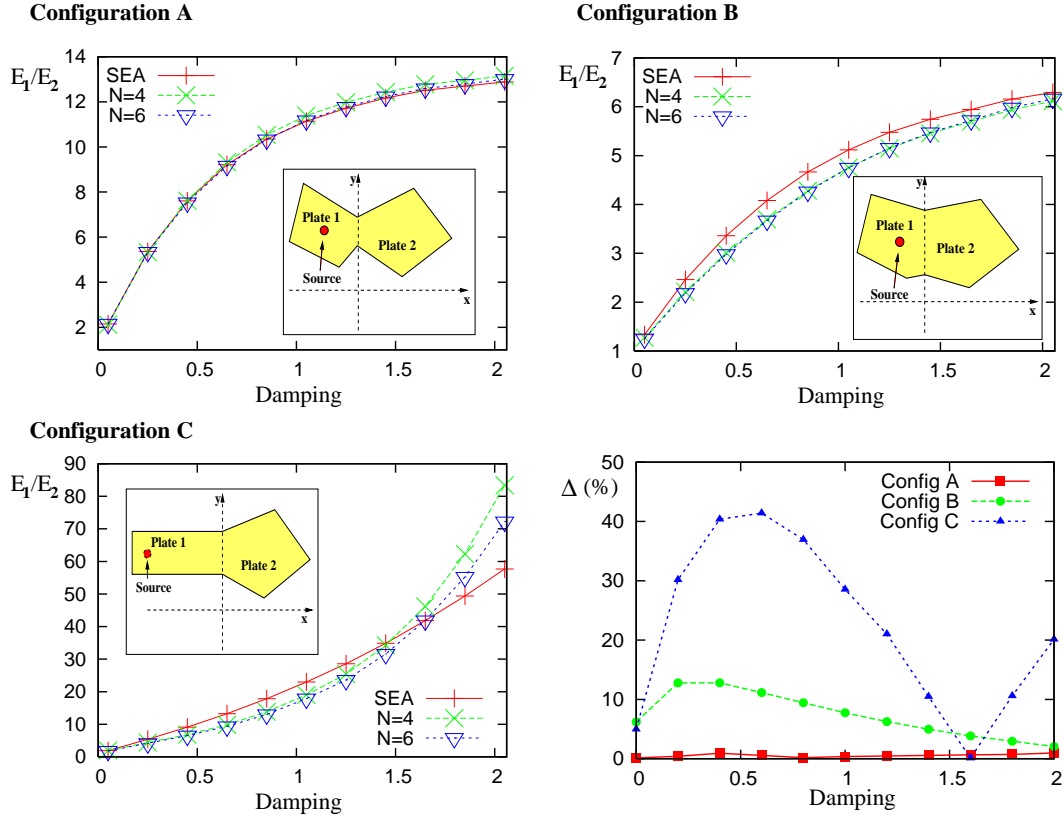}
\caption{Ratio of mean energy densities in plate 1 and 2 versus damping
coefficient $\alpha$ for the three different plate geometries; lower right
hand corner: relative difference (in percent) between results from 
SEA ($N=0$) and DEA with $N=6$ for the three configurations.
\label{fig:plate}
}
\end{figure*}

The plates considered in this study all consist of sets of straight boundaries 
\cite{note1}
-  such polygonal shapes 
are typical for many engineering applications.  Three different set-ups have 
been chosen (see Fig.\ \ref{fig:plate}): 
\begin{itemize}
\item{\bf configuration A} comprises two sub-systems of irregular shape 
with a line of intersection relatively small compared to the total length 
of the boundaries; the two subsystems are thus well separated and SEA is 
expected to work well. 
\item{\bf configuration B} consist of two plates where the line of 
intersection is of the order of the size of the system; the only dynamical 
barrier is posed by the BC itself. The standard SEA assumption of weak 
coupling and a quasi-stationary distributions in each subsystem may thus be 
violated.  (This configuration has also been studied in \cite{MR99,
Mac03}).
\item{\bf configuration C} has a left-hand plate with regular features and 
rays are channelled out of this plate effectively introducing long-range
correlations in the dynamics thus again violating a typical SEA assumption. 
In addition, the source is chooses at the far end of plate 1 in contrast
to the other two configurations with a source placed close to the intersection.
\end{itemize}
Note, that SEA results are in general insensitive to the position of the 
source, whereas actual trajectory calculations may well depend on the exact 
position especially for strong damping and for sources placed close to or 
far away from points of contact between subsections. 

Numerical calculations have been done for finite basis sets up to 
$n_1, n_2 = -N, \ldots N$ with $N \le 6$. This gives rise to matrices of the 
sizes $\mbox{dim} T = 2 (2N+1)^2$ with basis functions covering 
position and momentum coordinates uniformly in both subsystems. Energy
distributions have been studied as function of the damping rate $\alpha$. 
Note, that in the limit $\alpha \to 0$, the matrix $T$ has an 
eigenvalue one with eigenvector corresponding to an equidistributed energy 
density over both plates, see the discussion following Eq.\ (\ref{densi}). 
In the case of no damping, the ray dynamics explores the full phase space 
uniformly on the manifold $H(X) = \omega^2$ in the long time limit. 
Eq.\ (\ref{equb}) is singular for $\alpha = 0$ and the solutions 
become independent of the source distribution $\rho_0$ for $\alpha \to 0$.
One obtains
\[
\lim_{\alpha \to 0} \frac{\bar{\rho}_1}{\bar{\rho}_2} = 
\lim_{\alpha \to 0} \frac{\epsilon_1}{\epsilon_2} = 1
\]
where $\bar{\rho}_i$ denotes the mean ray density in plate $i$ averaged over 
the area of the plate and $\epsilon_i$ is the corresponding mean energy density
obtained from Eq.\ (\ref{wdens}).\\

\begin{figure*}
\includegraphics[width=0.8\textwidth]{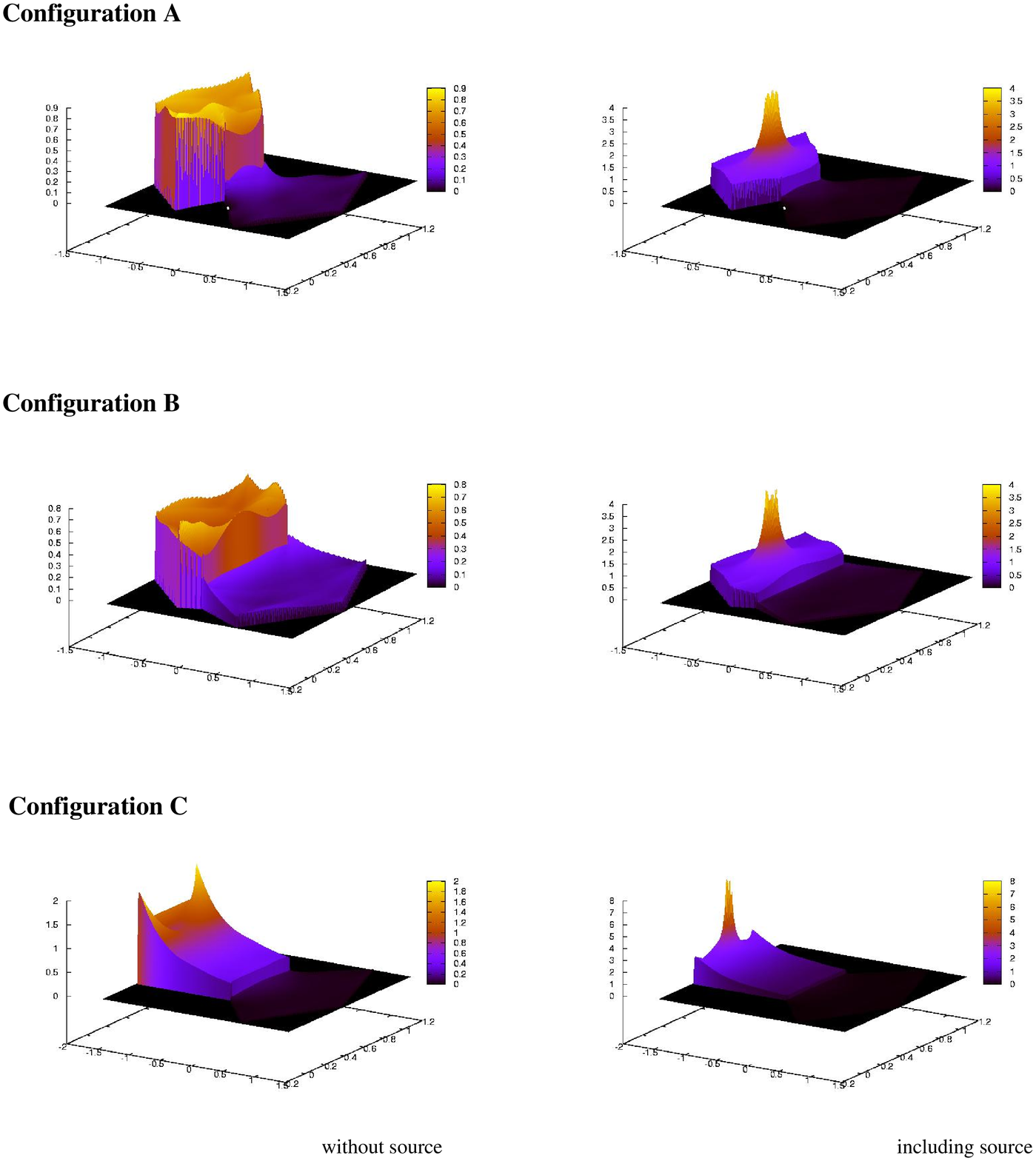}
\caption{Energy density distributions for the three configurations for 
$\alpha = 1$ and $N=6$; left: wave energy induced by reflections from the 
boundary; right: total wave energy distribution including direct contributions 
from the source. 
\label{fig:wf}
}  
\end{figure*}

Results for the relative energy density distribution for the two-plate
systems are shown in Fig.\ \ref{fig:plate}. Increasing the basis size - 
indicated here by the index $N$ - leads to fast convergence as is evident 
from the figures. An SEA-like treatment corresponds to $N=0$, here. The 
lower right hand panel in Fig.\ \ref{fig:plate} also shows the difference 
between an SEA and DEA treatment (the latter with $N=6$).

SEA works remarkably well for configuration A, for which the main SEA assumptions, 
that is, irregular shape,  well separated subsystems and
relative small damping, are fulfilled. The deviations between SEA and the 
high-resolution result $N=6$ are of the order of a few percent. Given that 
SEA describes the energy densities here in terms of a system of only 
two coupled equations, this clearly shows the power of SEA compared to, 
for example, ray tracing methods. In configuration B, 
the division into two subsystems is less clear-cut and deviations from 
SEA due to the strong coupling between the plates may be expected. Indeed, 
one finds a higher energy density in plate 2 than expected from SEA - 
energy dissipates into plate 2 before an equilibrium distribution is attained
in plate 1. The effect is here of the order of 
10 \% and thus still relatively small.

However, it is not too difficult to devise plate configurations where
significant deviations from SEA occur. In conf.\ C, plate 1 has 
rectangular shape thus acting as an effective channel for transporting
wave energy from plate 1 to plate 2; the plate shape thus induces long 
range correlations and memory effects into
the ray dynamics. In addition, the source has been placed away from the 
intersection magnifying both the influence of correlation effects as well as
short range effects due to absorption for large $\alpha$. One indeed finds
more wave energy in plate 2 than expected from an SEA treatment for 
small $\alpha$ - a clear sign that plate 1 acts as an effective channel. 
For large $\alpha$, however, the wave energy gets damped out before reaching 
the intersection due to the relative long path lengths caused by the 
position of the source. Thus, less wave energy than expected from an SEA 
treatment reaches plate 2 and the ratio $\epsilon_1/\epsilon_2$ is above the 
SEA curve. Note, that the deviations between DEA and SEA are now significant 
and in the region of about 50\%. \\

Within DEA, it is also possible to resolve the wave intensity distribution 
within each of the subsystems. The boundary distribution obtained from Eqs.\ 
(\ref{densb}) and (\ref{equb})can be mapped back into the interior using the operator 
${\cal L}_B^{-1}$ in each subsystems, see Eq.\ (\ref{densi}). The spatial 
resolution of the wave energy density contains 
important information about, for example, the (acoustic) radiation 
characteristics of sub-elements in the high frequency limit. In Fig.\
\ref{fig:wf}, typical intensity distributions are shown for the 
three plate configurations at a medium damping rate $\alpha = 1$ and $N=6$. 
The left
hand panel shows the wave distribution induced by rays reflected from the 
boundary (indirect contributions), the right hand panel also includes the 
direct rays emanating from the source point. (It is worth clarifying that 
only the indirect signals have been considered in the results presented in Fig.\
\ref{fig:plate}).  The wave intensity plots confirm the observations described 
earlier; while for configuration A and B, one can identify 
a quasi-equilibrium distribution in each subsystem characterised by a 
plateau-structure in each of the two plates, the correlated dynamics in 
configuration C leads to a smooth decay of the signal within 
plate 1. 

\section{Conclusions}
We have shown that ray tracing methods and SEA are closely related and 
that the latter is indeed an approximation of the former by smoothing out 
the details of ray dynamics within individual subsystems. We propose a 
numerical technique which interpolates between SEA and 
full ray tracing by resolving the ray dynamics on a  finer and finer 
scale. This is achieved by expressing the dynamics in terms of linear boundary
operators and representing those in terms of a set of basis functions on the 
boundary. The resolution of the dynamics is now determined by the number of 
boundary functions taken into account. 

We provide a derivation starting directly from a short wave length approximation 
of the wave equation and  leading all the way to setting up the basic 
DEA equations; we thus offer a step by step account of the approximations and 
simplifications made. The basic SEA assumptions can be tested systematically 
by relating them back to aspects of the ray dynamics.
Furthermore, extending SEA to DEA allows to  enhance the range of applicability 
of an SEA like treatment and will lead to robust and numerically efficient tools 
for determining energy density distributions in complex mechanical structures.\\

\noindent
\textbf{Acknowledgement:}

\noindent
The author would like to thank Oscar Bandtlow, Brian Mace and Anand Thite 
for stimulating discussions and Stewart McWilliam for providing some 
important references.  Support from the EPSRC through a 
{\em Springboard fellowship} is gratefully acknowledged.

\appendix
\section{Ray dynamics}
\label{app:eikonal}
A ray or classical dynamics associated with a wave equation (\ref{weqn}) can
be obtained via an Eikonal approximation writing the solutions in the form
of a phase $S(r)$ and amplitude $A(r)$;
assuming that the amplitude $A$ changes slowly on the scale of the wave length,
one obtains a governing equations for the phase $S$ alone. For example, for the
Helmholtz equation with $\hat{H} = c^2 \nabla^2$, one obtains
\begin{equation}
\label{eikonal}
c^2 (\nabla S)^2 = \omega^2\, ,
\end{equation}
where $c$ denotes the wave velocity (assumed to be constant here).
Dissipative terms are usually incorporated in the equation for the
amplitude $A$.  The Hamilton-Jacobi equation (\ref{eikonal})
can be solved by the method of characteristics.  After defining the
wave number vector $ p \equiv \nabla S$ (where we adopt the notation of
classical mechanics where $p$ refers to {\em momentum}) and the
Hamilton function
\begin{equation}
\label{Hamiltonian}
H({p},{r}) = c^2 {p}^2 = \omega^2\, ,
\end{equation}
one obtains the ray-trajectories $({ r}(\tau)), {p}(\tau))$ from
Hamilton's equations
\begin{equation}\label{eof-ray}
\dot{r} = \frac{d}{d\tau} { r} ={\nabla}_p H = 2\, c^2\, {p}; \quad
\dot{p} = \frac{d}{d\tau}{ p} = -{\nabla }_r H.
\end{equation}
The fictitious time $\tau$ is conjugated to the 'energy' $\omega^2$
and is related to the physical time by $t = 2 \omega \tau$.
The dimensionless {\em action} $S$ is given as
\begin{equation}
\label{action}
S({r},{r}_0) = \int_{r_o}^r d{r}' \, {p}({r}')
\end{equation}
where the integration is taken along a ray from ${r}_0$ to $r$ on the
manifold $H(r,p) = \omega^2$.
For homogeneous media ($c = constant)$, as considered here, on obtains
$S = |p| L$ with $L({r},{ r}_0)$, the length of the ray path from ${r}_0 \to
{ r}$.

The ray dynamics in mechanical structures consisting of coupled
sub-systems will typically entail reflection on boundaries,
partial reflection/transmission at interfaces between two media
and multi-component ray dynamics including mode conversion. The
latter may occur between pressure and shear ''rays'' at boundaries
for typical boundary conditions (such as free boundaries); note,
that the different wave components have different local wave velocities
and will thus follow different equations of motion (\ref{eof-ray}).

The number of different rays starting in ${r}_0$ (with arbitrary momentum)
and passing through $r$ increases (for fixed $\omega$) rapidly with the length
or the action of the ray trajectories. If the ray dynamics is
chaotic, that is, the ray solutions show exponential sensitivity to initial
conditions, one finds that the number of trajectories going from
${r}_0 \to {r}$ increases exponentially with their length
\cite{Gut90}. Regular dynamics
such as the solution of the Eq.\ (\ref{eof-ray}) for rectangular or
circular geometries leads to a power law increase in the number of ray
solutions from ${r}_0 \to  r$.

\section{Derivation of the ray-tracing Eq.\  (\ref{dens1})}
\label{app:ray-trace}
It will be shown here that Eq.\ (\ref{diagsum}) is equivalent to the ray tracing 
equations (\ref{dens1}), (\ref{dens2}). For further details on the  derivation,  see 
also \cite{Gut90}.  Starting point is Eq.\  (\ref{dens2}) 
(where we set $w \equiv 1$ here to simplify the notation), that is,   
\begin{equation} \label{dens21}
\rho({r}, {r_0}, \omega) = \int_0^{\infty} d\tau \int dp'\, 
\delta(r - \varphi_r^\tau(p',r_0))\, \delta(\omega^2 - H(p',r_0)) \, .
\end{equation}
We write the $\delta$-functions in the form
\begin{equation} \label{delta}
\delta(r - \varphi_r^\tau(p',r_0))\,\delta(\omega^2 - H(p',r_0)) =
\sum_j \frac{1}{D} \delta(\tau - \tau_j) \delta(p' - p'_j) 
\end{equation}
where the index $j$ counts all possible solutions of 
\[ \varphi_r^{\tau_j}(r_0,p_j') = r; \, H(p'^j,r_0) = \omega^2 \, .\]
These are the rays emanating from the source point $r_0$ and 
reaching the final point $r$ on the manifold $H = \omega^2$. The 
Jacobian $D$ is 
\[
D = \left|\frac{\partial (r, H)}{\partial (p',t)}\right| =  \left|\begin{array}{cc} 
\frac{\partial r}{\partial p'} &\frac{\partial H}{\partial p'}\\
 \frac{\partial r}{\partial t} &\frac{\partial H}{\partial t}\\                                   
\end{array}
\right| \, .
\]
Making use of the equation of motion (\ref{eof-ray}), one identifies 
${\partial H}/{\partial p'} = \dot{r}'$ and we have 
${\partial H}/{\partial t} = 0$. It is now convenient 
to switch to a local coordinate
systems $r = (r_\|, r_\perp); p = (p_\|, p_\perp)$ at the 
initial and final point where $r_\|$, $p_\|$ point along 
the trajectory in phase space. One obtains
\begin{equation} 
D = \left |\begin{array}{cccc } 
                        &                                               &           &\dot{r}_\|'\\
                        & \frac{\partial r}{\partial p'} &           &0\\
                        &                                               &           &\vdots\\
 \dot{r}_\|&0                                            &\cdots&0                                    
\end{array} \right|  
= |\dot{r}| |\dot{r'}| \left| \frac{\partial r_\perp}{\partial p_\perp}\right| \, 
= \left|A^{(g)}\right|^{-2}
\label{jac}
\end{equation}
where $A^{(g)}$ is the geometric contribution to the wave amplitude, 
Eq.\ (\ref{ag}).  Combining Eqs.\ (\ref{delta}) and (\ref{jac}) with 
(\ref{dens21}), one obtains the diagonal term (\ref{diagsum}).


\begin{thebibliography}{9999}
\bibitem{Gut90} Gutzwiller M C 1990 {\it Chaos in Classical and Quantum
                Mechanics} (Springer, New York)
\bibitem{Blu96} Bl\"umel R, Antonsen Jr. T M, Georgeot B, Ott E and 
Prange R E 1996 {\em Phys.\ Rev.\ Lett. }{\bf 76} 2467; 
Bl\"umel R, Antonsen Jr. T M, Georgeot B, Ott E, and Prange R E 1996 
{\em Phys. Rev.} E {\bf 53} 3284
\bibitem{BH98} Bogomolny E and Hugues E 1998 {\em Phys. Rev. A} {\bf 57} 5404
\bibitem{Cer01} Cerven\'y V 2001 {\em Seismic Ray Theory} (Cambridge 
University Press)
\bibitem{CI01} Chae K-S and Ih J-G 2001 {\em J.\ Sound Vib.} {\bf 240} 263
\bibitem{Cra96} Craik R J M 1996 {\em Sound transmission through
buildings: using statistical energy analysis} (Gower, Hampshire)

\bibitem{DasBuch}
Cvitanovi\'c P, Artuso R, Mainieri R, Tanner G and Vattay G 2008 
{\it Classical and Quantum Chaos}, {\tt www.nbi.dk/ChaosBook/}, 
Niels Bohr Institute, Copenhagen

\bibitem{CLK05} Cotoni V, Langley R S and Kidner M R F 2005 
{\em J.\ Sound Vib.} {\bf 288} 701


\bibitem{Fre97} Fred\"o C R 1997 {\em J.\ Sound Vib.} {\bf 199} 645

\bibitem{Gla89} Glasser A S (Ed) 1989 {\em An Introduction to Ray Tracing}
(Academic Press)

\bibitem{GV79} Grammaticos B and Voros A 1979 
{\it Ann. of Physics \bf 123} 359


\bibitem{GMW98} Guhr T, M\"uller-Groeling A and Weidenm\"uller H A 1998
                {\em Phys.\ Rep.} {\bf 299} 189

\bibitem{Haa01} Haake F 2001{\em 
Quantum Signatures of Chaos}, (2nd edn., Springer, Berlin)

\bibitem{Her94} Heron K H 1994 {\em Phil.\ Trans.\ R.\ Soc.\ London A} 
{\bf 346} 501; reprinted in \cite{KP94} 

\bibitem{KP94} Keane A J and Price W G 1994 {\em Statistical Energy 
Analysis}, Cambridge University Press, Cambridge.

\bibitem{Kut00} Kuttruff H 2000 {\em Room Acoustics}, 4th edition, Spon 
Press, London

\bibitem{Lan90} Langley R S 1990 {\em J.\ Sound Vib.} {\bf 141} 207
\bibitem{Lan92} Langley R S 1992 {\em J.\ Sound Vib.} {\bf 159} 483; 
see also Langley R S and Bercin A N 1994 in \cite{KP94} p 59 

\bibitem{LB99} Langley R S and Bremner P 1999 {\em J.\ Acoust.\ Soc.\ Am.} 
{\bf 105} 1657

\bibitem{LB04} Langley R S and Brown A W M 2004 {\em J.\ Sound Vib.} 
{\bf 275} 823; {\em ibd} {\bf 275} 847

\bibitem{LC04} Langley R S and Cotoni V 2004 {\em J.\ Acoust.\ Soc.\ Am.} 
{\bf 115} 706

\bibitem{LB98} Le Bot A 1998 {\em J.\ Sound Vib.} {\bf 211} 537;
Le Bot A 2002 {\em J.\ Sound Vib.} {\bf 250} 247;
Le Bot A 2007 {\em J.\ Sound Vib.} {\bf 300} 763

\bibitem{Lyo67} Lyon R H 1967 {\em J.\ Acoust.\ Soc.\ Am.} {\bf 45} 545

\bibitem{LD95} Lyon R H and DeJong R G 1995 {\em Theory and Application 
of statistical energy analysis} (2nd edn.\ Boston, MA: Butterworth-Heinemann)

\bibitem{MH91} Mc Kown J W and Hamilton Jr R L 1991 {\em Ray Tracing as a 
Design Tool for Radio Networks}, IEEE Network Magazine 27

\bibitem{MR99} Mace B R and Rosenberger J 1999 {J.\ Sound Vib.} {\bf 212} 
395

\bibitem{MS00} Mace B R and Shorter P J 2000 {J.\ Sound Vib.} {\bf 233} 
369

\bibitem{Mac03} Mace B R 2003 {J.\ Sound Vib.} {\bf 264} 391; 
 Mace B R 2005 {J.\ Sound Vib.} {\bf 279} 141 

\bibitem{Ric00} Richter K 2000 {\em Semiclassical Theory of Mesoscopic 
               Quantum Systems}, (Springer, Berlin)

\bibitem{SL05} Shorter P J and Langley R S 2005 {\em J.\ Sound Vib.} 
{\bf 288} 669

\bibitem{ST02}
S\o ndergaard N and Tanner G 2002  {\em Phys.\ Rev.\ E} {\bf 66} 066211

\bibitem{TS07a}
Tanner G and S\o ndergaard N 2007a {\em Wave chaos in acoustics and 
elasticity} {\em J.\ Phys.\ A} {\bf 40} R443

\bibitem{TS07b} Tanner G and Soendergaard N 2007b 
{\em Phys.\ Rev.\ E} {\bf 75} 036607

\bibitem{Son07} S\o ndergaard N 2007 {\it J.\ Phys.\ A} {\bf 40} 5067

\bibitem{Sto99} St\"ockmann H-J 1999 {\em Quantum Chaos - an introduction}
(CUP, Cambridge)

\bibitem{Wea82} Weaver R L 1982 {\em J.\ Acoust.\ Soc.\ Am.\ \bf 71}
1608

\bibitem{Wea89} Weaver R L 1989, {J.\ Sound Vib.} {\bf 130} 487
\bibitem{note1} {To be precise, polygonal shapes as
considered here lead to pseudo-integrable dynamics which is strictly speaking
not even ergodic; at the level of approximation considered here, the decay
of correlation in the dynamics of irregular polygons is sufficient in
principle to test the SEA assumptions.} 
\end{thebibliography}
\end{document}